\documentclass[journal=nalefd,manuscript=letter,layout=twocolumn]{achemso}

\usepackage{chemformula} % Formula subscripts using \ch{}
\usepackage[T1]{fontenc} % Use modern font encodings

\usepackage{units}
\usepackage{upgreek}
\usepackage{color}

\author{Sascha Kolatschek}
\email{s.kolatschek@ihfg.uni-stuttgart.de}
\altaffiliation{These authors contributed equally to this work}
\author{Cornelius Nawrath}
\altaffiliation{These authors contributed equally to this work}
\author{Stephanie Bauer}
\altaffiliation{These authors contributed equally to this work}
\author{Jiasheng Huang}
\author{Julius Fischer}
\author{Robert Sittig}
\author{Michael Jetter}
\author{Simone L. Portalupi}
\author{Peter Michler}
\affiliation[Stuttgart University]
{Institut f\"ur Halbleiteroptik und Funktionelle Grenzfl\"achen (IHFG), Center for Integrated Quantum Science and Technology (IQ$^{ST}$) and SCoPE, University of Stuttgart, Allmandring 3, 70569 Stuttgart, Germany}

\title[An \textsf{achemso} demo]
  {Bright Purcell enhanced single-photon source in the telecom O-band based on a quantum dot in a circular Bragg grating}


\begin{document}

\begin{abstract}
The combination of semiconductor quantum dots (QDs) with photonic cavities is a promising way to realize non-classical light sources with state-of-the-art performances in terms of brightness, indistinguishability and repetition rate.
In the present work we demonstrate the coupling of an InGaAs/GaAs QDs emitting in the telecom O-band to a circular Bragg grating cavity. We demonstrate a broadband geometric extraction efficiency enhancement by investigating two emission lines under above-band excitation, inside and detuned from the cavity mode, respectively. In the first case, a Purcell enhancement of 4 is attained. For the latter case, an end-to-end brightness of \unit[1.4]{\%} with a brightness at the first lens of \unit[23]{\%} is achieved. 
Using p-shell pumping, a combination of high count rate with pure single-photon emission (g$^{(2)}$(0) = 0.01 in saturation) is achieved. Finally a good single-photon purity (g$^{(2)}$(0) = 0.13) together with a high detector count rate of \unit[191]{kcps} is demonstrated for a temperature of up to \unit[77]{K}.
\end{abstract}

In the emerging fields of quantum communication and quantum cryptography one main requirement is the use of bright non-classical light sources. In recent years semiconductor quantum dots (QDs) have become of great significance in the field of quantum optics \cite{Michler2017}, due to their outstanding photon emission properties in terms of single-photon emission \cite{Schweickert2018}, indistinguishability \cite{He2013} and (pair) entanglement fidelity \cite{Huber2018}. Although state-of-the-art performances have been achieved in the near-infrared region \cite{Ding2016, Somaschi2016, Dousse2010, Huber2017, Chen2018,Zhai2021}, the versatility of QDs also allows for single-photon emission in the telecom O-band and C-band both in the InAs/InP \cite{Takemoto2004,Miyazawa2005,Benyoucef2013,Muller2018} and InGaAs/GaAs material system \cite{Paul2015,Dusanowski2017,Ward2005,Paul2017}.
Especially in long-distance applications employing optical fibers, sources in the telecom wavelength regime are desirable due to both, low photon absorption and wave-packet dispersion.
Furthermore, using the existing fiber network, single-photons in the telecom O-band are favorable not only because of zero wave-packet dispersion but also for wavelength division multiplexing with classical communication signals in the telecom C-band \cite{Xiang2020}.
One of the main challenges for QD single-photon sources is the high refractive index contrast between the surrounding semiconductor material and the air. In the near-infrared, numerous approaches with cavities like micropillars \cite{Somaschi2016,Wang2016} and photonic crystal cavities \cite{Englund2005,Hennessy2007} have been realized to enhance the extraction efficiency next to geometric approaches such as microlenses \cite{Sartison2017,Gschrey2015}, photonic trumpets \cite{Munsch2013} and nanowires \cite{Claudon2010}.

Recently, circular Bragg gratings \cite{Davanco2011,Sapienza2015} have been given particular attention due to a new proposed flip-chip layout \cite{Yao2018}, avoiding a fragile free standing membrane \cite{Kolatschek2019}. This allowed for the demonstration of entangled photon pair emission with high brightness and indistinguishability for GaAs QDs emitting around \unit[780]{nm} \cite{Liu2019} as well as In(Ga)As QDs at \unit[900]{nm} \cite{Wang2019}. 

However, for the telecom regime, most of the results have been demonstrated in the InAs/InP material system \cite{Birowosuto2012,Kim2016,Kors2017,Haffouz2018,Jaffal2019,Lee2021} with approaches in the InGaAs/GaAs material system being limited to etched microlenses and mesas \cite{Sartison2018,Srocka2020}, dielectric antennas \cite{Yang2020} and micropillars \cite{Chen2017}. 

The feasibility of circular Bragg gratings in the telecom regime was suggested with simulations \cite{Rickert2019} as a promising approach, however, the realization of such a structure was still outstanding.

\begin{figure}
	\includegraphics[width=\linewidth]{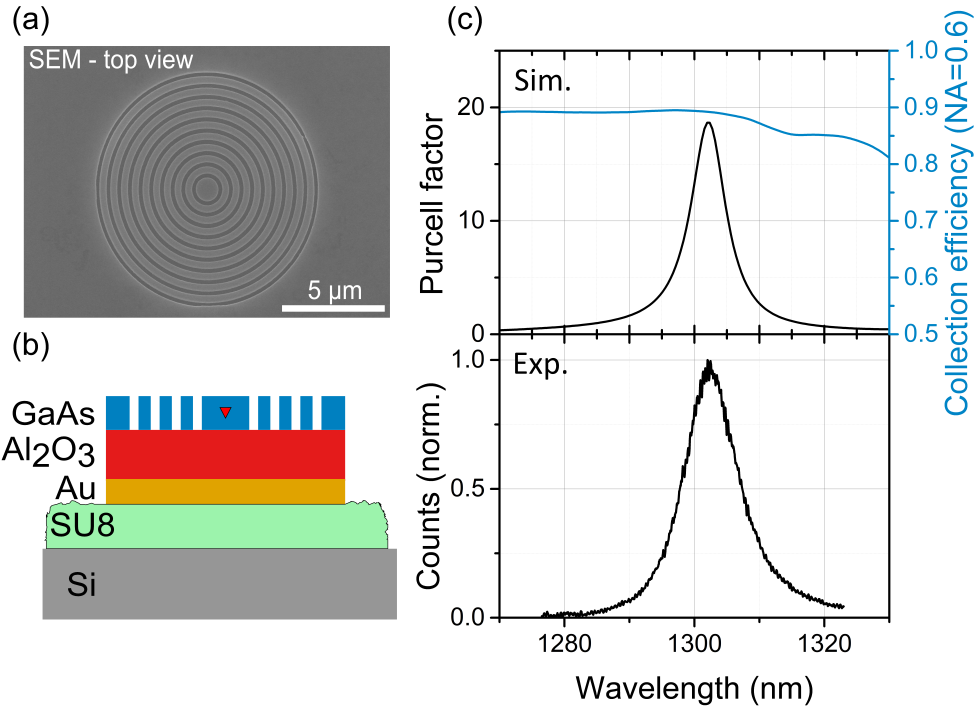}%
	\caption{(a) SEM top view of a processed circular Bragg grating. (b) Schematic cross section of the cavity with an exemplary self-assembled QD. (c) Top: Simulated Purcell factor and collection efficiency into NA=0.6. Bottom: Cavity mode measured in $\upmu$-PL under strong cw above-band pumping.\label{fig:1}}
\end{figure}

Here, we investigate emission from self-assembled InGaAs/GaAs QDs in a circular Bragg grating operating in the telecom O-band (Figure \ref{fig:1}).
Coupling between emitter and cavity is achieved, yielding a Purcell enhancement by a factor of around 4. Furthermore, bright single-photon emission is shown for transitions under above-band and p-shell pumping. Finally, investigations confirm the preservation of bright single-photon emission for elevated temperatures up to \unit[77]{K}.

\begin{figure*}
	\includegraphics[width=\linewidth]{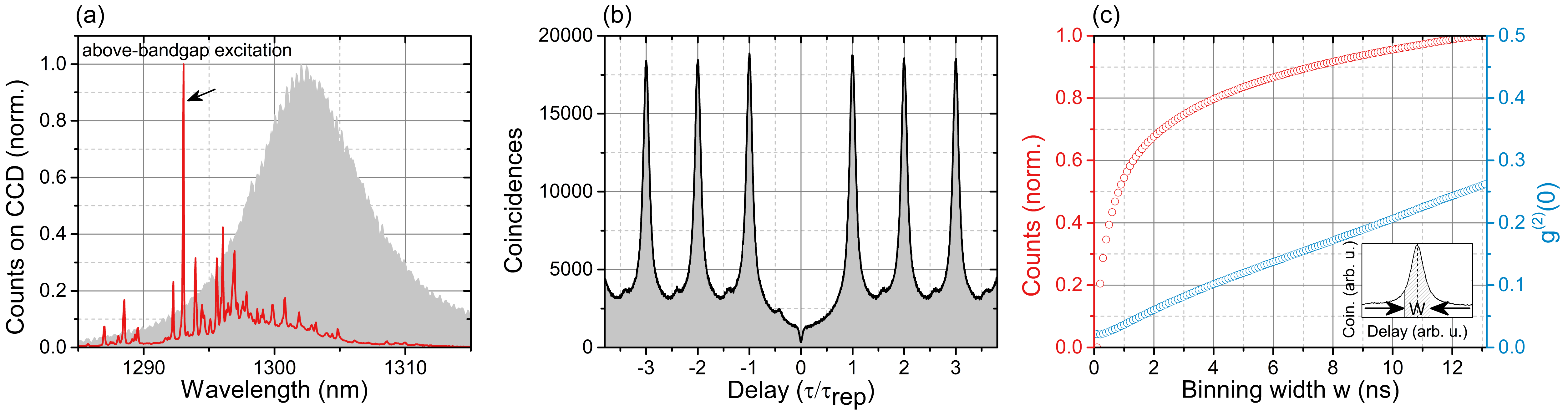}%
	\caption{(a) Normalized $\mu$-PL spectra under above-band pumping (cavity position as shaded area) with the investigated emission line marked by the black arrow. (b) Pulsed g$^{(2)}$($\tau$) measurement integrated over \unit[5]{minutes} and displayed with a binning of \unit[50]{ps}. (c) Evaluation of the g$^{(2)}$(0) and count rate for different binning widths w. The binning width w is hereby defined by the width symmetrically around the peak at each repetition were the counts are taken into account (inset). \label{fig:2}}
\end{figure*}

For optimal light extraction, a flip-chip process together with a gold backside mirror is chosen as proposed by Yao and co-authors \cite{Yao2018}. In contrast to previous realizations of this cavity at lower wavelengths \cite{Liu2019,Wang2019}, here a \unit[320]{nm} thick Al$_{2}$O$_{3}$ is used as the spacer layer between the GaAs membrane and the gold mirror (Figure \ref{fig:1}b). Thereby, the surface quality increased, minimizing scattering of the light at defects.

To optimize the device design, finite-difference time-domain (FDTD) simulations are performed. With the investigated cavity, a theoretical Purcell enhancement of 18.6 combined with a Q-factor of around 200 can be expected (Figure \ref{fig:1}c top). Unique for this cavity design is the broadband collection with an efficiency of over \unit[80]{\%} into an NA of 0.6 for a wavelength range of more than \unit[60]{nm}. Furthermore misplacing the emitter from the center by up to \unit[400]{nm} has no effect on the high collection efficiency, however, strongly reduces the Purcell factor (not shown).
Hence, this cavity structure exhibits high tolerance to spatial and spectral mismatch while maintaining bright single-photon emission.
The measured cavity mode under strong continuous-wave (cw) above-band pumping shown in the bottom of Figure \ref{fig:1}c and the simulated cavity mode are in good agreement. The strong pumping allows for a uniform feeding of the cavity mode via background emission. Due to a small ellipticity of the structure, the fundamental cavity mode exhibits a polarization splitting of \unit[1]{nm}. Small fabrication imperfections, e.g. not perfectly vertical sidewalls, lead to a measured Q-factor of around 130 in comparison to a simulated Q-factor under ideal conditions of 200. Taking the measured Q-factor into account, the maximally achievable Purcell factor for perfect spatial and spectral overlap is reduced to 12.

For the optical characterization of a non-deterministically placed cavity with at least 2 QDs in the central disk, a confocal micro-photoluminescence ($\upmu$-PL) setup is used with the sample mounted inside a helium flow cryostat and cooled down to \unit[4]{K}. To excite the QD under non-resonant excitation above the bandgap of the barrier material (above-band excitation), a \unit[775]{nm} pulsed laser with a repetition rate of \unit[76]{MHz} and a pulse width of \unit[3]{ps} is used. A bright excitonic line can be seen at \unit[1293]{nm}, around \unit[9]{nm} blue-shifted with respect to the cavity mode (Figure \ref{fig:2}a).

To investigate the single-photon purity of the emission, the excitonic line is spectrally filtered (full width at half maximum (FWHM) = \unit[0.04]{nm}) and send to a fiber-based Hanbury-Brown and Twiss setup reaching a total count rate of up to \unit[1.3]{Mcps}. A second-order autocorrelation is integrated for \unit[5]{minutes} and shown in Figure \ref{fig:2}b.  
At zero time delay, a clear antibunching dip can be observed. In order to evaluate the g$^{(2)}$(0), a binning width as large as the repetition is applied to account for all the measured counts on the detector. Comparing the area around zero time delay to the mean area of the outer peaks (for this measurement, no blinking of the emission line was observed) yields a g$^{(2)}$(0) = 0.262 $\pm$ 0.001. 
Due to a relatively long decay time, an overlap between the neighboring peaks can be observed. The small peaks in between the laser repetition are a measurement artifact due to electronic reflections in the setup.
Correcting the measured count rate for the non-zero g$^{(2)}$(0) results in a true single-photon count rate of \unit[1.1]{Mcps} on the detectors, corresponding to an end-to-end brightness of \unit[1.4]{\%} (the correction is analogues to Ref.\citenum{Lee2021}). Considering the setup efficiency of \unit[6.4]{\%}, a brightness at the first lens of \unit[23]{\%} is achieved. Here, it is worth noting that due to the applied above-band pumping scheme multiple transitions are excited. Therefore the stated brightness gives only a lower bound as no competing transitions e.g. trions of the same QD, are taken into account \cite{Nowak2014}.
To assess the performance of the cavity for time-gated applications, different binning widths w are applied for the g$^{(2)}$(0) evaluation (Figure \ref{fig:2}c). The count rate is hereby directly proportional to the square root of the mean area under the outer peaks integrated over the binning width while g$^{(2)}$(0) shows a linear dependence with the binning width. Therefore by time-gating, an improved g$^{(2)}$(0) value can be achieved with only a minor decrease  in the count rate, e.g. an increase of the single-photon purity by a factor of two can be gained at the cost of only \unit[15]{\%} in the count rate.

Due to the layout of the cavity, the central disk is located on an Al$_{2}$O$_{3}$ layer, completely isolated from the surrounding GaAs. As a consequence, this may lead to a more sensitive charge carrier environment, e.g. after strongly pumping the cavity mode a change in the relative probability of different QD transitions due to changes in the charge dynamics and surface states can be observed under above-band pumping.
In between the measurements, the sample was stored for some days outside the cryostat under a nitrogen atmosphere. This resulted in a substantial permanent change of the emission probabilities.
\begin{figure}
	\includegraphics[width=\linewidth]{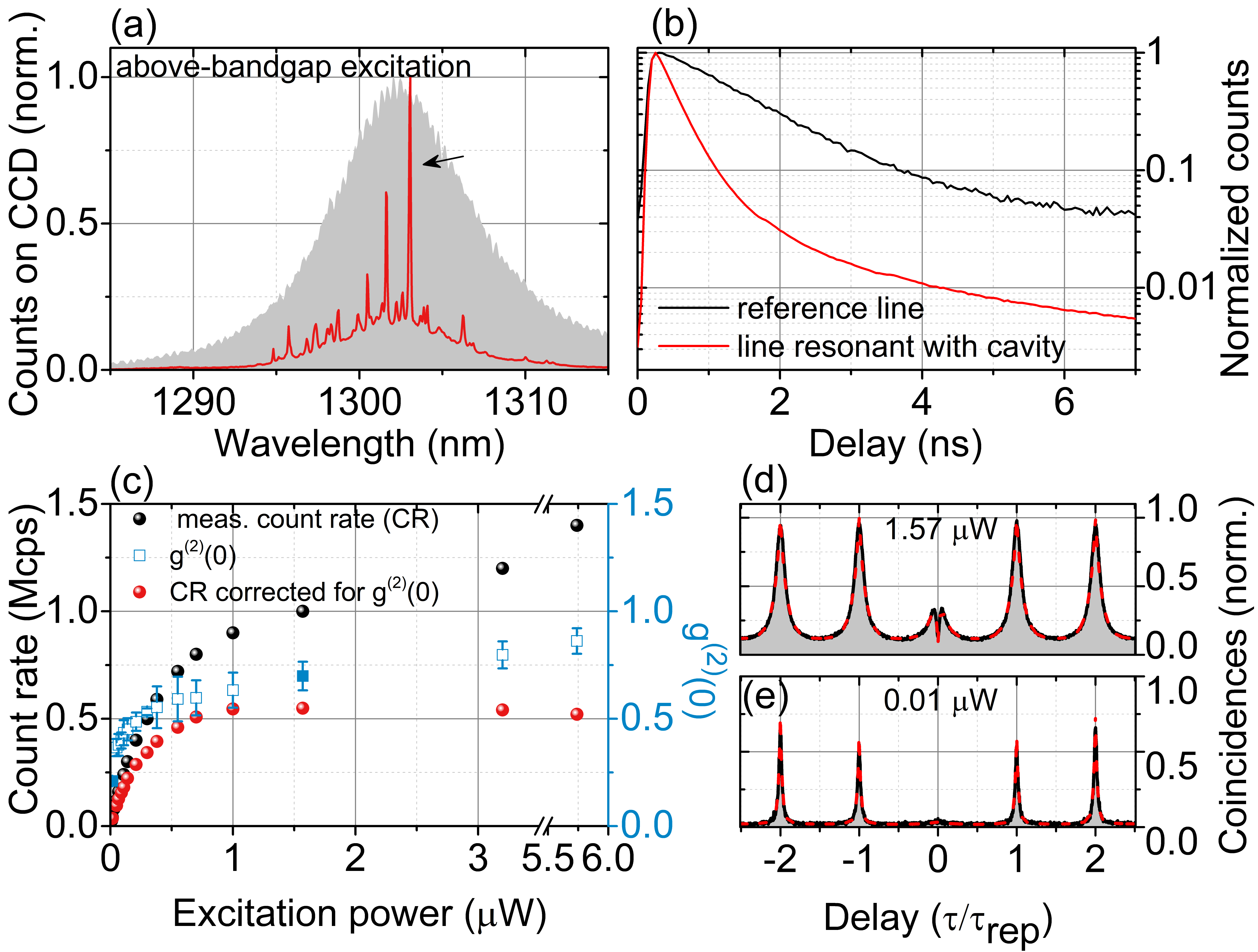}%
	\caption{(a) Normalized $\mu$-PL spectra under above-band pumping (cavity position as shaded area) with the investigated emission line marked by the black arrow. (b) Decay time measurements of the emission line inside the cavity and an exemplary reference line in the membrane outside the cavity. (c) Measured count rate (black) as a function of the excitation power and corrected for the non-perfect g$^{(2)}$(0) values (red) as well as the corresponding g$^{(2)}$(0) values (blue squares). The filled squares denote exemplary g$^{(2)}$($\tau$) for saturation power (d) and low power (e). \label{fig:3}}
\end{figure}

While the excitonic line shown in Figure \ref{fig:2}a required stronger pumping than before, a second excitonic emission line from a different QD almost resonant with the cavity mode emerged for low pumping powers (Figure \ref{fig:3}a). For this line, only a small spectral mismatch with regard to the cavity mode can be seen. A conclusive way to probe the spatial overlap between the QD and the cavity mode is by measuring the Purcell enhancement of the transition. The decay time measurement of the considered emission line together with an exemplary reference line is shown in Figure \ref{fig:3}b. With respect to measurements of decay times from multiple QDs outside the cavity ($\tau_{mean}=\unit[1209.4]{ps}$ with a standard deviation of $\unit[226.3]{ps}$) a strongly reduced decay time of $\tau_{Purcell}=\unit[300.8\pm 0.8]{ps}$ is observed. By comparing the decay times, a Purcell enhancement of 4.0 $\pm$ 0.7 is deduced. The deviation from the optimal theoretical Purcell factor of 12 is attributed mainly to the non-perfect spatial overlap between emitter and cavity mode due to the non-deterministic approach.
The measured Purcell enhancement not only improves the photon indistinguishability but could in the future also allow for higher repetition rates \cite{Anderson2020}.
In order to evaluate the count rate of the emission line a power series is performed (Figure \ref{fig:3}c black circles). However, no clear saturation behavior is observed. 
Therefore, in a next step power dependent second-order correlation measurements are carried out (Figure \ref{fig:3}c blue squares). For fitting, the central cw-dip as well as blinking are taken into account in the form of an envelope function \cite{Santori2001,Miyazawa2016}. The good agreement with the measured data given by the fitting model is verified for high and low excitation powers as shown in Figure \ref{fig:3}d and \ref{fig:3}e respectively.
In saturation, the evaluated g$^{(2)}$(0) = 0.70 $\pm$ 0.07 shows that a strong multi-photon probability is present.
At high excitation power, a strong refilling of the QD, presumably from nearby charge carrier traps can be observed (Figure \ref{fig:3}d).
However, the cw g$^{(2)}$(0) value can be estimated by the depth of the central dip yielding g$^{(2)}_{cw}$(0) = 0.10 $\pm$ 0.04. 
This shows that the measured counts stem from one emitter and the g$^{(2)}$(0) is mainly deteriorated by the strong refilling. To deduce the true single-photon count rate on the detector, the measured count rate is corrected for the non-perfect g$^{(2)}$(0) yielding at saturation a count rate of \unit[550]{kcps}.

\begin{figure}
	\includegraphics[width=\linewidth]{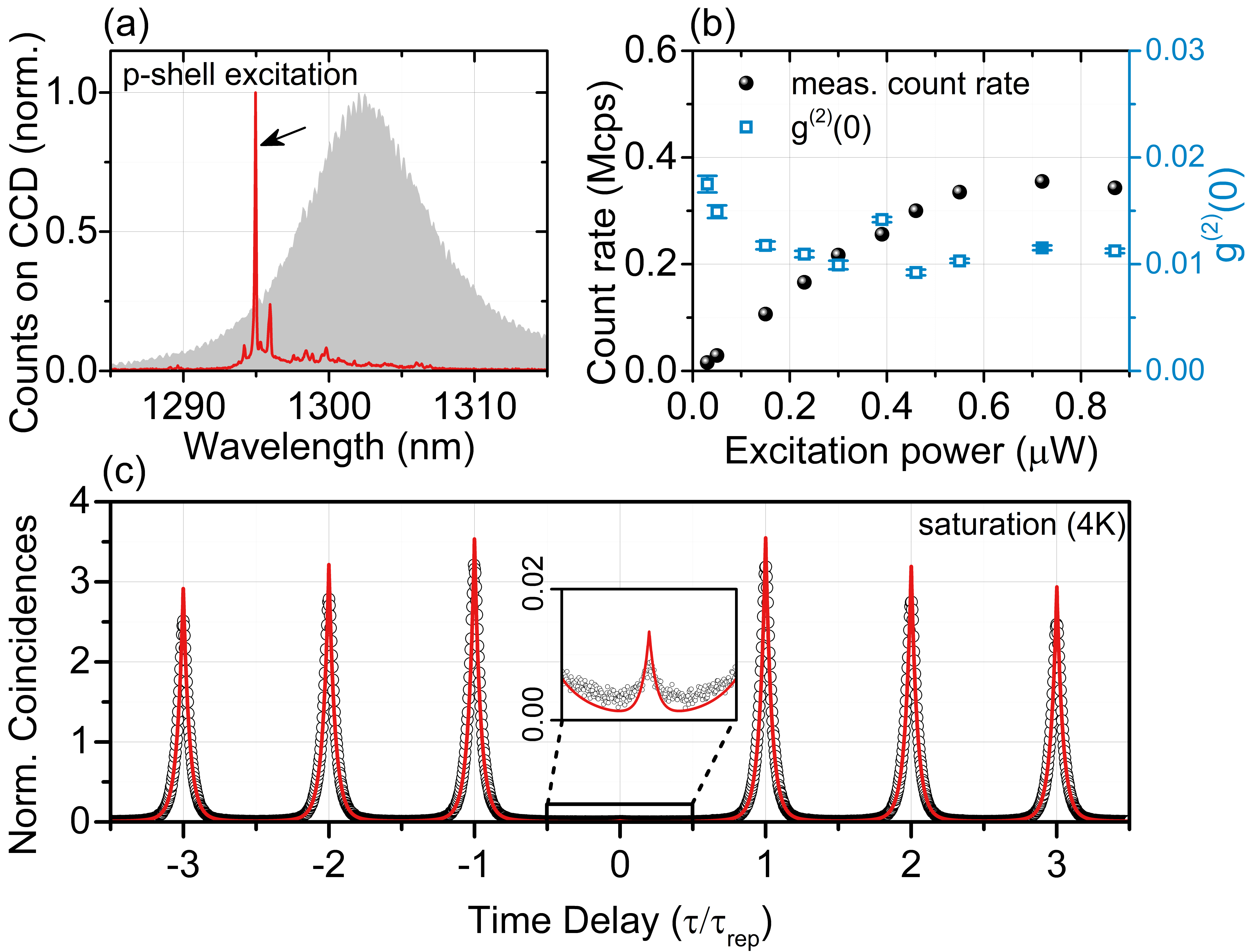}%
	\caption{(a) Normalized $\mu$-PL spectra under p-shell pumping (cavity position as shaded area) with the investigated emission line marked by the black arrow. (b) Measured count rate on the detector (black) and power-dependent g$^{(2)}$(0) values (blue squared). The filled square denotes the value g$^{(2)}$(0) = 0.0115 $\pm$ 0.0002, obtained for saturation (c) g$^{(2)}$($\tau$) for saturation power.\label{fig:4}}
\end{figure}

In order to avoid refilling, a quasi-resonant excitation scheme via the discrete energy state (p-shell) is applied \cite{Dusanowski2017,Lee2021}. Around \unit[1200]{nm}, p-shell resonances of multiple transitions can be found. In the following, an exemplary line with its p-shell at \unit[1207]{nm} is investigated. This allows for an easy filtering of the excitation laser with off-the-shelf \unit[1250]{nm} long-pass filters.
With the excitation laser at \unit[1207]{nm}, this bright emission line, again blue-shifted with respect to the cavity mode, is observed (Figure \ref{fig:4}(a)). 
In contrast to the previous two investigated emission lines, here no fine structure splitting is observed hinting strongly towards the presence of a charged transition. 
To assess the brightness and purity of this emission line, a power series in combination with power dependent second-order correlation measurements is performed. In saturation a count rate of \unit[355]{kcps} is detected (Figure \ref{fig:4}b). The fitted g$^{(2)}$(0) values are mostly power independent and range between 0.01 and 0.02 with a value of 0.012 under saturation power. The small increase for low excitation powers can probably be attributed to a slightly different background behavior for high and low excitation powers. 
Further measurements on different excitonic as well as charged transitions show similar g$^{(2)}$(0) values.
The excellent g$^{(2)}$(0) values prove that indeed by adapting the pumping scheme, a strong suppression of the refilling mechanism can be achieved. This is also reflected in the fitting function, where the amplitude of the secondary decay is strongly reduced with respect to the above-band excitation case.

Furthermore, a strong blinking of the emitter on a timescale of around \unit[100]{ns} is observed as seen in Figure \ref{fig:4}c which is taken into account when normalizing the g$^{(2)}$($\tau$) measurements. The resulting long off-time of the transition significantly limits the maximal attainable count rate to around \unit[25]{\%} in comparison to the case without blinking. 

Final measurements show the relevance of the described structure in real-world implementations. Indeed, for practical applications, it is advantageous to achieve bright single-photon emission at elevated temperatures making the source compatible with cryocoolers or liquid nitrogen. Therefore the temperature is increased to \unit[40]{K} and \unit[77]{K} respectively and the charged transition seen in Figure \ref{fig:4}a is investigated under saturation power in terms of count rate and single-photon purity. 
As higher temperatures lead to a phonon broadening of the emission line, the spectral selection window is increased to FWHM = \unit[0.1]{nm} to avoid spectral filtering with a width smaller than the zero-phonon line \cite{Olbrich2017}. 
To verify that this is not the case for the approach for the previous measurements at \unit[4]{K} with the smaller spectral selection window of \unit[0.04]{nm}, the corresponding measurement are repeated using the larger spectral selection window. The absence of a significant degradation regarding the detected count rate as well as g$^{(2)}$(0) confirms that at \unit[4]{K} the FWHM of the emission line is much smaller than the spectral selection window.
\begin{figure}
	\includegraphics[width=\linewidth]{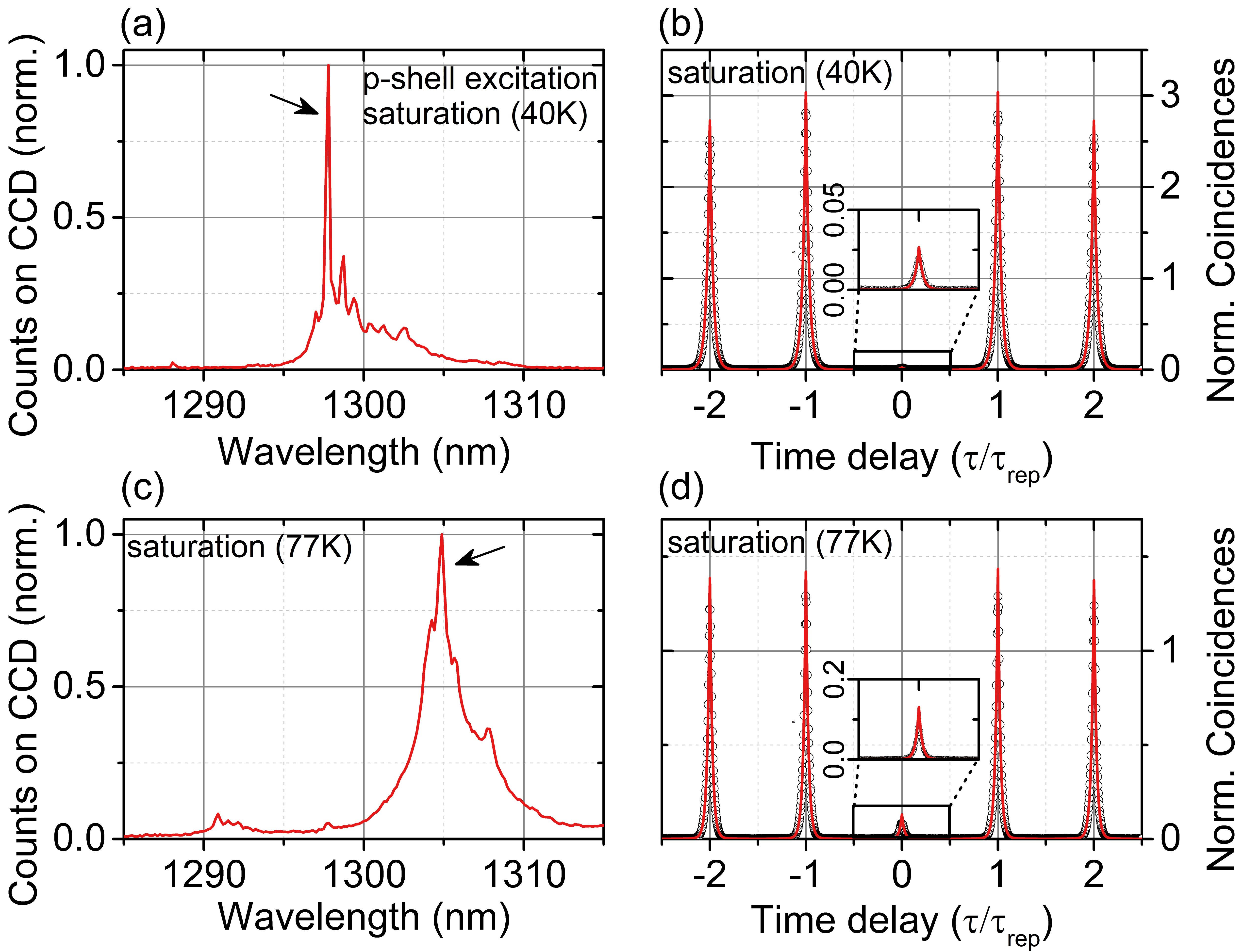}%
	\caption{(a) Normalized $\mu$-PL spectra under p-shell pumping at (a) \unit[40]{K} and (c) \unit[77]{K} with the investigated emission line marked by the black arrow. g$^{(2)}$($\tau$) in saturation power at (b) \unit[40]{K} with g$^{(2)}$(0) = 0.026 $\pm$ 0.001  and (d) \unit[77]{K} with g$^{(2)}$(0) = 0.127 $\pm$ 0.004.\label{fig:5}}
\end{figure}
The spectrum obtained at \unit[40]{K} is shown in Figure \ref{fig:5}a. An expected red-shift of the emission line of around \unit[3]{nm} is observed \cite{Olbrich2017}.
With a combined count rate of \unit[262]{kcps} on the detectors, a g$^{(2)}$(0) of 0.026 $\pm$ 0.001 is obtained (Figure \ref{fig:5}b). 
Further increasing the temperature to \unit[77]{K} results in an additional red-shift and line broadening (Figure \ref{fig:5}c). Here a decrease of the count rate in saturation to \unit[191]{kcps} on the detectors is observed, which corresponds to around \unit[50]{\%} of the count rate at \unit[4]{K}.
Despite the relatively high temperature, a g$^{(2)}$(0) of 0.127 $\pm$ 0.004 is measured. This value, clearly below 0.5, proves the good single-photon purity of the investigated emission line inside the circular Bragg grating even at liquid-nitrogen temperature.
Furthermore, the reported g$^{(2)}$(0) values shows, that almost no deterioration of the single-photon purity is present up to \unit[40]{K}. The small deviation at \unit[77]{K} can most probably be attributed to background emission from neighboring transitions as suggested in Figure \ref{fig:5}c, spectrally overlapping with the investigated transition.
Comparing Figure \ref{fig:5}b and \ref{fig:5}d, a decrease of the FWHM of the peaks with increasing temperature is seen. This points towards an increase in non-radiative charge carrier loss channels which results in a decrease of the count rate as observed for higher temperatures.

In conclusion, we realized a circular Bragg grating in the telecom O-band with bright transition lines exhibiting single-photon emission, both spectrally resonant and blue-shifted with respect to the cavity mode. Using above-band excitation a true single-photon count rate with an end-to-end brightness of \unit[1.4]{\%} and first-lens brightness of \unit[23]{\%} was achieved in combination with a g$^{(2)}$(0) of 0.262 $\pm$ 0.001, clearly below 0.5. A decay time reduction of 4.0 $\pm$ 0.7 due to the Purcell effect was observed for a bright line spectrally resonant with the cavity mode. This will in the future allow for higher repetition rates further increasing the amount of usable photons. However, despite a pronounced central dip, strong refilling prevented even better g$^{(2)}$(0) values. Therefore a p-shell pumping scheme was applied. This allowed for a strong reduction of the refilling processes resulting in saturation in a g$^{(2)}$(0) of 0.012 combined with a count rate of \unit[355]{kcps} on the detector. In the future, by applying electric fields to stabilize the charge environment, the observed blinking could be reduced, enabling even higher count rates. Finally, single-photon emission with strong multi-photon suppression was shown up to \unit[77]{K} (g$^{(2)}$(0) = 0.127). The capability of emitting on-demand single photons even at elevated temperature, i.e. 40K and 77K, without the need for complex schemes to filter the excitation laser light is very appealing for future technological implementations. Indeed, the achieved results prove that such kind of telecom QD-cavity structures can be operated in combination with compact cryocoolers (which typically can reach temperatures around 40K) or even in nitrogen based processes. This would allow for a drastic reduction of the overall setup footprint and power consumption making the realized sources appealing for real-world implementations as quantum information and satellite based quantum cryptography.

\begin{acknowledgement}

The authors gratefully acknowledge the funding by the German Federal Ministry of Education and Research (BMBF) via the project Q.Link.X (No. 16KIS0862) and the European Union's Horizon 2020 research and innovation program under Grant Agreement No. 899814 (Qurope). The work reported in this paper was partially funded by Project No. EMPIR 17FUN06 SIQUST and Project No. EMPIR 20FUN05 SEQUME. This project has received funding from the EMPIR programme co-financed by the Participating States and from the European Union's Horizon 2020 research and innovation program.

\end{acknowledgement}

\bibliography{references}

\end{document}